\documentclass{aa} 
\usepackage{txfonts}
\usepackage[latin1]{inputenc}  
\usepackage{graphicx}

\begin{document}
\title{Destruction of formic acid by soft X-rays in star-forming regions}
%
\author{H. M. Boechat-Roberty\inst{1} \and  S. Pilling\inst{1}\inst{2} \and A. C. F. Santos\inst{3}}
\institute{Observatório do Valongo, Universidade Federal do Rio de
Janeiro, Ladeira Pedro Antônio 43, CEP 20080-090, Rio de Janeiro,
RJ, Brazil. \and Instituto de Química, Universidade Federal do Rio
de Janeiro, Ilha do Fundão, CEP 21949-900, Rio de Janeiro, RJ,
Brazil \and Instituto de Física, Universidade Federal do Rio de
Janeiro, Caixa Postal 68528, CEP 21941-972, Rio de Janeiro, RJ,
Brazil.}
\offprints{H.M. Boechat-Roberty, \email{heloisa@ov.ufrj.br}}
\date{Received / Accepted}
%
\abstract{Formic acid is much more abundant in the solid state, both
in interstellar ices and cometary ices, than in the interstellar gas
(ice/gas $\sim 10^{4}$) and this point remains a puzzle. The goal of
this work is to experimentally study ionization and
photodissociation processes of HCOOH (formic acid), a glycine
precursor molecule. The measurements were taken at the Brazilian
Synchrotron Light Laboratory (LNLS), employing soft X-ray photons
from toroidal grating monochromator TGM) beamline (200 - 310 eV).
Mass spectra were obtained using photoelectron photoion coincidence
(PEPICO) method. Kinetic energy distributions and abundances for
each ionic fragment have been obtained from the analysis of the
corresponding peak shapes in the mass spectra. Photoionization and
photodissociation cross sections were also determined. Due to the
large photodissociation cross section of HCOOH it is possible that
in PDRs regions, just after molecules evaporation from the grain
surface, formic acid molecules are almost totally destroyed by soft
X-rays, justifying the observed low abundance of HCOOH in the
gaseous phase. The preferential path for the glycine formation from
formic acid may be through the ice phase reaction. }
\titlerunning{Destruction of Formic Acid by Soft X-Rays in Star Forming Regions}
\authorrunning{Boechat-Roberty, Pilling \& Santos}
\maketitle

\section{Introduction}
Formic acid (HCOOH) has been observed in various astronomical
sources: protostellar ices NGC 7538:IRS9 (Ehrenfreund \& Shutte,
2000), comet Hale-Bopp (Crovisier \& Bockelée-Morvan 1999), dark
molecular clouds (Ohshi et al. 1992) and Photodissociation Regions
(PDRs) associated with Hot Molecular Cores (Zuckermann et al. 1971,
Liu et al. 2001, Liu et al. 2002). Cazaux et al. (2003) observed an
extremely rich organic inventory in the hot core around the low-mass
protostar IRAS 16293-24 with abundant amounts of molecules such as
formic acid, acetaldehyde (CH$_3$CHO), methyl formate (CH$_3$OCHO),
dimethyl ether (CH$_3$OCH$_3$) and acetic acid (CH$_3$COOH).
Bottinelli et al. (2004) also reported the detection of formic acid
in the hot core of the low mass protostar NGC133-IRAS4A. It is known
that protostars are sources of X-rays (Koyama et al. 1996).

The simplest amino acid, glycine, (NH$_2$CH$_2$COOH), was recently
detected in the molecular clouds SgrB2, Orion KL and W51 (Kuan et
al. 2003, 2004). In these objects the precursor molecules like
ammonia, formic acid and acetic acid have already been observed
(Turner 1991 and Sutton et al. 1985). Liu et al. (2002) pointed out
the importance of performing studies on formic acid since it is the
simplest organic molecule and shares common structural elements with
biologically important species such as acetic acid and glycine.

Sgr B2, Orion KL and W51 are massive star-forming regions where the
presence of widespread UV and X-ray fields could trigger the
formation of photodissociation regions (PDRs). X-ray photons are
capable of traversing large column densities of gas before being
absorbed. X-ray-dominated regions (XDRs) in the interface between
the ionized gas and the self-shielded neutral layers could influence
the selective heating of the molecular gas. The complexity of the
region possibly allows a combination of different scenarios and
excitation mechanisms to coexist within the complex (Goicoechea et
al. 2004).

The formation of HCOOH in molecular clouds by gas-phase reactions
has been suggested by Irvine et al. (1990). On the other hand,
experiments and models of grain-surface chemistry suggest that HCOOH
is also readily produced in icy grain mantles (Tielens \& Hagen
1982; Allamandola \& Sandford 1990; Charnley 1995). Upon mantle
evaporation, HCOOH will be released into the gas phase and can be
detected at millimeter wavelengths (Liu et al. 2001, 2002).
Ehrenfreund et al. (2001) have shown that formic acid is more
abundant in the solid state, both in interstellar ices and cometary
ices, than in the interstellar gas (ice/gas $\sim 10^{4}$); this
point remains a puzzle and more laboratory work is necessary to
clarify this question.

\begin{figure*}
 \centering
 \includegraphics[width=18cm]{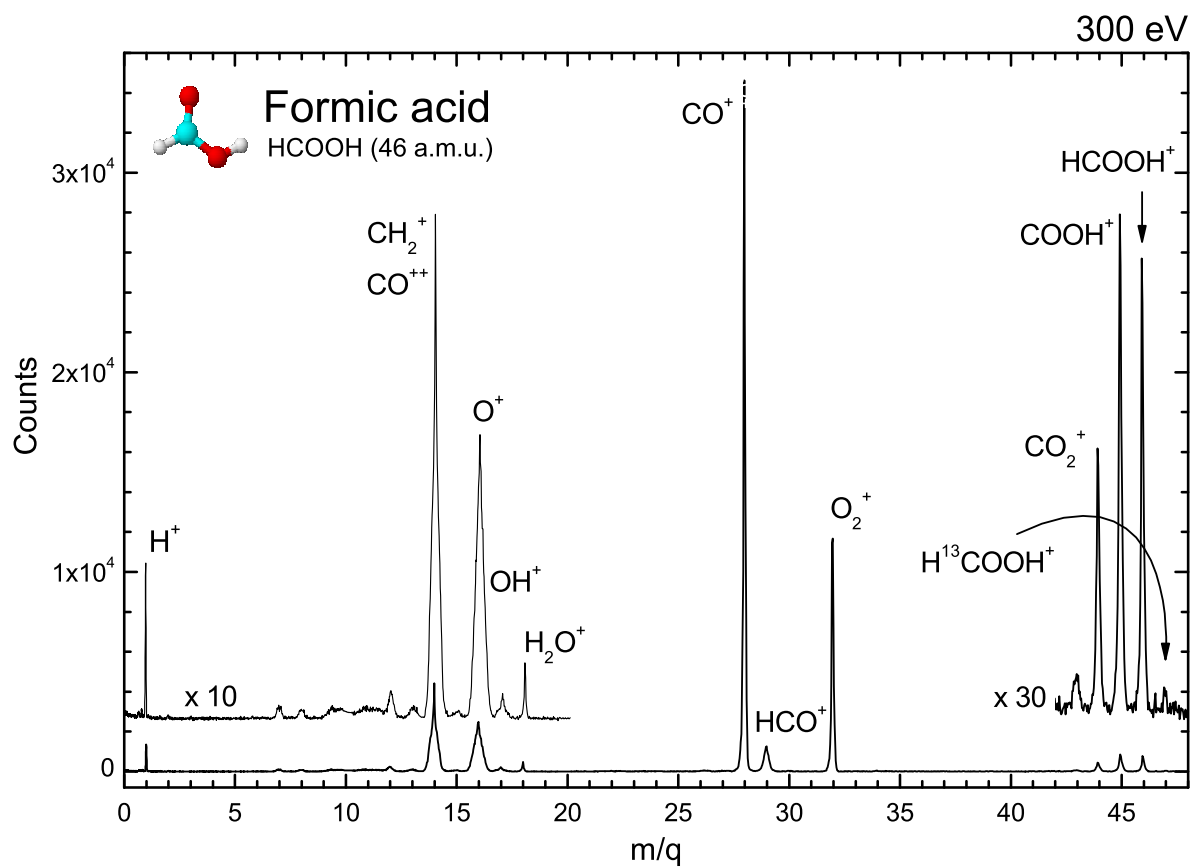}
 \caption{Time-of-flight mass spectrum of HCOOH molecule recorded at 300 eV.}
 \label{fig-ms}
\end{figure*}

Sorrell (2001) has proposed the formation of the carboxyl group
(COOH) as a stable radical in the UV photolysis of H$_2$O/CO ice.
Reactions in the gas phase and in the ice phase (``ice-solvated
phase'') have been considered by Woon (2002) via quantum chemical
modeling. Mendoza et al. (2004) have shown by theoretical
calculations that COOH can also be generated on graphitic surfaces
via adsorbate-adsorbate reactions.

The photodissociation of formic acid has been studied experimentally
and theoretically in the vacuum ultraviolet (VUV) region (Su et al.
2000; Tabayashi et al. 1999; Schwell et al. 2002). However the
results cannot explain the ice/gas ratio, since about 20 \% of
formic acid survives the UV radiation. Despite some photo absorption
studies in the X-ray range (Ishii \& Hicthcock 1987; Prince et al.
2003) there are no studies focusing on the photodestruction of
molecules by soft X-rays. The question about the relative abundances
of formic acid in solid and gas phase is still open (Ehrenfreund et
al. 2001). The present work aims to examine the photodissociation of
formic acid by soft X-rays, at energies around the carbon K edge.

\section{Experimental}

The experimental set-up has been described in detail elsewhere (Lago
et al. 2004 and de Souza et al. 2000). Briefly, the experiment was
performed at the Brazilian Synchrotron Light Source (LNLS),
Campinas, S\~ao Paulo, Brazil. Soft X-rays photons ($\sim10^{12}$
photons/s) from a toroidal grating monochromator (TGM) beamline
(100-310 eV) perpendicularly intersect the effusive gaseous sample
inside a high vacuum chamber. The base pressure in the vacuum
chamber was in the $10^{-8}$ Torr range. During the experiment the
chamber pressure is maintained below $10^{-5}$ Torr. The gas needle
is kept at ground potential. The emergent photon beam is recorded by
a light sensitive diode. The sample was commercially obtained from
Sigma-Aldrich with purity better than 99.5\%. No further
purification was used except for degassing the liquid sample by
multiple freeze-pump-thaw cycles before admitting the vapor into the
chamber. Conventional time-of-flight mass spectra were obtained
using the correlation between one Photoelectron and a Photoion
Coincidence (PEPICO). The ionized recoil fragments produced by the
interaction with the photon beam are accelerated by a two-stage
electric field and detected by two micro-channel plate detectors in
a chevron configuration after being mass-to-charge (m/q) analyzed by
a time-of-flight mass spectrometer (TOF-MS). They produce up to
three stop signals to a time-to-digital converter (TDC) started by
the signal from one of the electrons accelerated in the opposite
direction and recorded without energy analysis by two micro-channel
plate detectors. Besides PEPICO spectra, other two kinds of
coincidence mass spectra were obtained simultaneously (Pilling et
al. 2005, in prep.): PE2PICO spectra (PhotoElectron Photoion
Photoion Coincidence) and PE3PICO spectra (PhotoElectron Photoion
Photoion Photoion Coincidence). Those spectra have ions coming from
double and triple ionization process, respectively, that arrive
coincidentally with a photoelectron. In both cases the multiple
ionization is associated with the Auger process. Of all signals
received by the detectors only about 10\% come from PE2PICO and 1\%
from PE3PICO spectra, reflecting that the majority contribution is
indeed due to single event coincidence.

The pressure at the interaction region (volume defined by the gas
beam and the photon beam intersection) was estimated to be $\sim$ 1
Torr. At this pressure about 30\% of HCOOH molecules are dimers
(Halford 1942, Barton et al. 1969) but no obvious consequences of
these clusters were seen in the detected spectra. The measurements
were done at room temperature.

The first stage of the electric field (708 V/ cm) consists of a
plate-grid system crossed by the light beam at the center. The
TOF-MS was designed to have a maximized efficiency for ions with
energies up to 30 eV (Willey and McLaren 1955). The secondary
electrons produced in the ionization region are focused by an
electrostatic lens polarizing the electron grid with 800 V, designed
to focus them at the center of the micro-channel plate detector.
Negative ions may also be produced and detected, but the
corresponding cross-sections are negligible. Our experimental setup
does not have an electron energy analyzer to measure the
photoelectron energies.

\section{Results and discussion}

Figure~\ref{fig-ms} shows a mass spectrum of the formic acid
obtained at 300 eV photon energy. Some general observations can be
made. The CO$^{+}$ ion production is the most likely outcome
(36-40\%) in the dissociation of the formic acid molecule in the
energy range studied in the present work. This is followed by
CH$_2^+$ (with a possible contribution of same m/q ion CO$^{++}$)
(18-27 \%), O$^+$, HCO$^+$ (or COH$^+$) and O$_2^+$ ($\sim$ 14 \%)
fragments, the latter coming from rearrangements. The CH$_2^+$ +
CO$^{++}$ peak rises to very broad structures usually associated
with a large intrinsic kinetic energy.

\begin{figure}
\resizebox{\hsize}{!}{\includegraphics{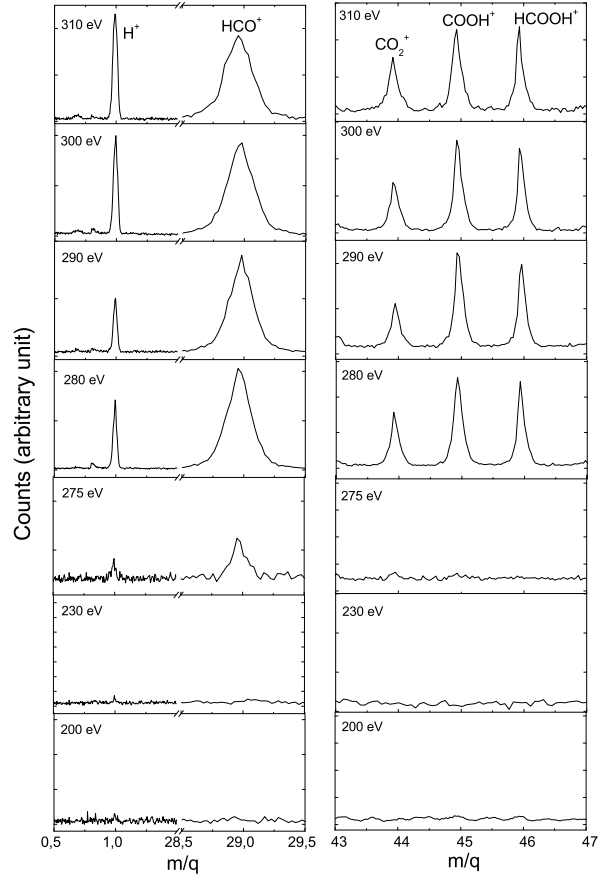}}
\caption{Time-of-flight mass spectra of formic acid molecule showing
details of H$^+$, HCO$^+$ (left panel) and CO$_2^+$, COOH$^+$ and
HCOOH$^+$ ions (right panel) for several photon energies.}
\label{fig-zoom}
\end{figure}

\begin{table*}
\centering \caption{Relative intensities (partial ion yield - PIY)
and kinetic energy $U_0$ release by fragments in the formic acid
mass spectra, as a function of photon energy. Only fragments with
intensity $>$ 0.1 \% were tabulated. The estimated experimental
error was 10\%}. \label{tab-piy}
\begin{tabular}{ l l l r r r r r r r }
\hline \hline
\multicolumn{2}{c}{Fragments}    &  & \multicolumn{7}{c}{PIY (\%) / $U_0$ (eV)}\\
\cline{1-2}  \cline{4-10}
   $m/q$        & Attribution    &  & 200 eV       & 230 eV       & 275 eV       & 280 eV       & 290 eV       & 300 eV       & 310 eV       \\
\hline
1       & H$^+$                  &  &  -           & 0.71 / 0.24  & 0.19 / 0.24  & 1.33 / 2.17  & 1.32 / 2.17  & 2.36 / 2.87  & 2.83 / 4.80  \\
7       & CH$_2^{++}$; CO$^{++++} ? $ &  &  -      & 1.57 / 30.5  & 0.86 / 11.0  & 0.72 / 13.8  & 0.47 / 8.84  & 0.52 / 9.97  & 0.64 / 10.0  \\
8       & O$^{++}$               &  &  -           & -            & 0.65 / 2.45  & 0.57 / 12.0  & 0.30 / 10.9  & 0.38 / 8.71  & 0.49 / 2.45  \\
9.3     & CO$^{+++}$ ?           &  & -            & -            & 1.21 / 52.0  & 0.91 / 87.8  & 0.83 / 67.5  & 0.91 / 33.5  & 0.95 / 45.8  \\
12      & C$^+$                  &  &  -           & -            & 0.26 / 1.28  & 0.56 / 0.50  & 0.49 / 0.40  & 0.93 / 0.98  & 1.19 / 0.98  \\
13      & CH$^+$                 &  &  -           & -            &  -           & 0.30 / 1.18  & 0.30 / 1.04  & 0.49 / 1.04  & 0.47 / 0.16  \\
14      & CH$_2^+$; CO$^{++}$    &  & 26.7 / 0.07  & 24.6 / 0.11  & 22.6 / 0.72  & 20.6 / 0.72  & 17.7 / 0.61  & 17.1 / 0.43  & 18.2 / 0.96  \\
16      & O$^+$                  &  & 20.9 / 0.30  & 20.0 / 1.99  & 16.8 / 1.99  & 15.9 / 1.65  & 13.1 / 0.96  & 13.4 / 1.22  & 14.9 / 1.50  \\
17      & OH$^+$                 &  & -            & -            & -            & 0.45 / 0.06  & 0.49 / 0.17  & 0.59 / 0.13  & 0.75 / 0.17  \\
18      & H$_2$O$^+$               &  & -            & 0.66 / 0.12  & 0.25 / 0.05  & 0.55 / 0.05  & 0.67 / 0.03  & 0.64 / 0.03  & 0.60 / 0.05  \\
28      & CO$^+$                 &  & 37.9 / 0.02  & 37.4 / 0.02  & 39.9 / 0.02  & 37.7 / 0.02  & 41.1 / 0.02  & 39.2 / 0.02  & 37.3 / 0.03  \\
29      & HCO$^+$                &  & -            & 0.64 / 0.10  & 0.31 / 0.03  & 2.89 / 0.13  & 3.08 / 0.13  & 3.24 / 0.17  & 3.19 / 0.16  \\
32      & O$_2^+$                &  & 14.4 / 0.03  & 13.4 / 0.03  & 14.1 / 0.03  & 13.2 / 0.02  & 15.5 / 0.02  & 14.4 / 0.02  & 13.3 / 0.01  \\
44      & CO$_2^+$               &  & -            & -            & 0.16 / 0.03  & 0.63 / 0.02  & 0.56 / 0.02  & 0.72 / 0.02  & 0.75 / 0.03  \\
45      & COOH$^+$               &  & -            & 0.76 / 0.01  & 0.14 / 0.16  & 1.18 / 0.03  & 1.26 / 0.03  & 1.29 / 0.02  & 1.19 / 0.03  \\
46      & HCOOH$^+$              &  & -            & 0.14 / 0.11  & 0.11 / 0.06  & 0.91 / 0.01  & 0.97 / 0.01  & 1.02 / 0.02  & 0.94 / 0.01  \\
\hline \hline
\end{tabular}
\end{table*}

In Figure~\ref{fig-zoom} we show details of three regions of a mass
spectrum of formic acid obtained in the energy range from 200 to 310
eV, around the C 1s resonance ($1s \rightarrow 2\pi^{\ast})$,
exhibiting peak profiles of H$^+$, HCO$^+$, COO$^+$, COOH$^+$ and
HCOOH$^+$ ions. To clarify the mechanism leading to the molecular
fragmentation, we observe that the mentioned ions appear in the mass
spectra only above 275 eV suggesting that the main contribution to
the fragmentation processes comes from a resonant-type Auger decay,
which takes place after the electron excitation to unoccupied
Rydberg or antibonding orbitals, and might be responsible for the
appearance of the HCOOH$^+$ parent ion. This kind of resonant-type
Auger decay creates a hole in the valence orbital and its final
state is identical to a direct valence photoionization. Below 275
eV, the hole can be in an inner valence orbital, favoring the full
molecular fragmentation preferentially into the CO$^+$ plus neutrals
or CH$_2^+$ plus neutrals.

The reactive ions CO$^+$ and COH$^+$ that are photodissociation
product, were detected toward the H II region Monoceros R2 (Rizzo
2003) and in three photodissociation region, the reflection nebula
NGC7023, the Orion Bar and the planetary nebula NGC7027 (Fuente et
al. 2003). Usero et al. (2004) reported the first extragalactic
detection of the reactive ion COH$^+$ in the circumnuclear disk of
Active Galactic Nuclei, NGC 1068. They concluded that X-rays can
heavily influence the physical conditions and chemical abundances of
molecular gas and that the circumnuclear disk has become a giant
X-ray Dominated Region (XDR).

Another feature is the fate of the H$^+$, C$^+$ and CH$^+$ ions
below 280 eV photon energy. One possible explanation for this
behavior is the fact that dissociation occurs less often. It seems
to be associated with the fate of the parent molecule. One
possibility is that the molecule remains intact and neutral. Only
radiative decay could result in uncharged molecules, but the
radiative yields are expected to be very low in the present energy
range. Corresponding to the increase in the H$^+$, C$^+$ and CH$^+$
yields is a relative decrease in the relative yields of CH$_2^+$ +
CO$^{++}$, suggesting that the peak at $m/q$=14 is dominated by the
CH$_2^+$ and not by the doubly charged CO$^{++}$ fragment. From the
mass spectra one can observe an increase in the H$^+$ peak intensity
due to the increase in the kinetic energy release as a function of
the photon energy (2.2 eV at 280 and 290 eV, 3.9 at 300 eV and 4.8
at 310 eV). Below 280 eV, the kinetic energy of the H$^+$ fragment
is not enough to escape, being recaptured to form the CH$_2^+$ ion.
As the photon energy increases, so does the kinetic energy of H$^+$,
increasing the probability of forming H$^+$.

\begin{figure}
\resizebox{\hsize}{!}{\includegraphics{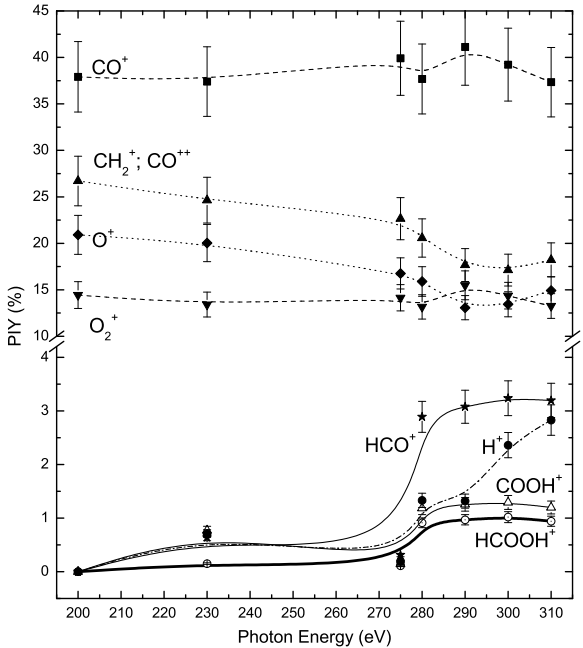}}
\caption{Partial ion yield (PIY) of some PEPICO fragments of HCOOH
molecule as a function of photon energy.} \label{fig-piy}
\end{figure}

The doubly charged fragments O$^{++}$ and CH$_2^{++}$ are also
observed from 275 eV and above. The absence of more doubly ionized
fragments in the PEPICO spectra indicates that the doubly ionized
formic acid dissociates preferentially via charge separation. The
dynamics of the doubly ionized formic acid molecule will be the
subject of a future publication.

Figure~\ref{fig-piy} shows the partial ion yields (PIY) for the most
significant outcomes in the dissociation of formic acid in the
200-310 eV photon energy range, the yields of H$^+$, HCO$^+$,
COOH$^+$ and HCOOH$^+$ are also shown. A small dip can be seen in
the fractions of CO$^+$ and O$_2^+$ at the C1s edge. Above 290 eV
this fraction shows a drop as a function of energy. The fraction of
CH$_2^+$ fragments also drops around the C 1s edge while the
fraction of the C$^+$, CH$^+$, HCO$^+$, OH$^+$ and H$^+$ exhibit a
gradual increase, being totally absent below 275 eV. This
observation strongly indicates that H$^+$, C$^+$, CH$^+$,
H$_2$O$^+$, OH$^+$, and HCO$^+$ are only formed after the
normal-type Auger decay (a small contribution of HCO$^+$ of roughly
0.3 \% is observed at 275 eV).

\begin{figure}
\resizebox{\hsize}{!}{\includegraphics{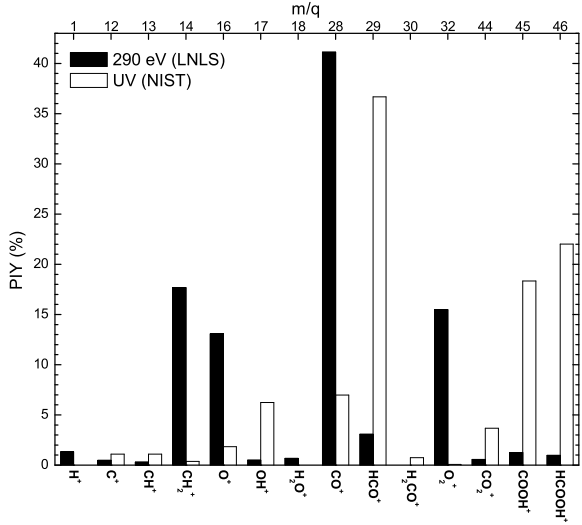}} \caption
{Comparison between partial ion yield (PIY) in soft X ray field and
UV field.} \label{fig-piycomp}
\end{figure}

In Figure \ref{fig-piycomp} we show a comparison between partial ion
yield in soft X-rays (290 eV) obtained at LNLS with the one in UV
(by 70 eV electrons) \footnote {The effect of a 70 eV electron is
very similar to 21.21 eV (He I Lamp) photons; in both the main
ionization occurs in a valence shell (see discussion in Lago et al.
2004)} from NIST \footnote {National Institute of Standards and
Technology http://webbook.nist.gov/chemistry/}. The molecular ion
HCOOH$^+$ is more destroyed by soft X-rays than by a low UV field.
The partial ion yields of several fragments also present
\begin{figure}
\resizebox{\hsize}{!}{\includegraphics{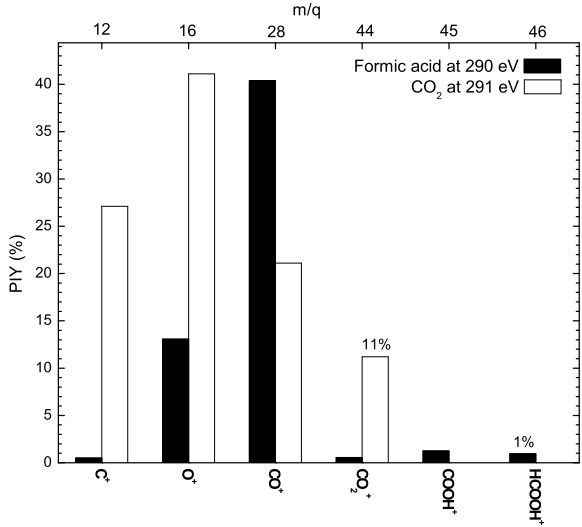}} \caption
{Comparison between partial ion yield (PIY) of HCOOH (this work) and
CO$_2$ (Hitchcock et al. 1979).} \label{fig-piycompCO2}
\end{figure}
differences in the X-ray field when compared with UV field, for
example the large enhancement of CO$^+$ production by X-rays. The
same is true for O$^+$, O$_2^+$ and CH$_2^+$ fragments. The opposite
behavior occurs with HCO$^+$ and OH$^+$ which seem to be more
efficiently produced by UV photons, as has been previously reported
(Suto etal 1988, Su et al. 2000; Schwel et al. 2002).

A comparison of partial ion yield results of the HCOOH and CO$_{2}$
molecules measured at photon energies of 290 and 291 eV,
respectively, can be seen in Figure~\ref{fig-piycompCO2}. The
CO$_{2}$ survives about 10 \% more than HCOOH and its main products
by photodissociation are O$^+$ and C$^+$ at 40 \% and 27 \%,
respectively. Those ionic fragments represent only 13 \% and 1 \% of
all fragments in the case of formic acid photodissociation. The
HCOOH dissociation is two times more efficient for CO$^+$ production
than CO$_2$ dissociation.

From the observations described in the preceding paragraphs, we
suggest that below the C 1s edge, the fragmentation of the formic
acid molecule is dominated by the pathways
\\\\
\begin{tabular}{r c l}
\centering
HCOOH + $h\nu$ & $\longrightarrow$ & HCOOH$^+ + e^-$ \\
HCOOH$^+$      & $\longrightarrow$ & CO$^+$ + H$_2$O (or H + OH)  \\
HCOOH$^+$      & $\longrightarrow$ & CH$_2^+$ + O$_2$ (or O + O) \\
HCOOH$^+$      & $\longrightarrow$ & O$^+$ + HCOH (?) \\
HCOOH$^+$      & $\longrightarrow$ & O$_2^+$ + CH$_2$ (or CH + H) \\
HCOOH$^+$      & $\longrightarrow$ & HCO$^+$ + OH (or O + H) \\
\end{tabular}

\subsection{Kinetic energy release (heating) of the ionic fragments}

Several authors have recently focused on the pathway of formation of
biomolecules present in the star-forming region and other
gaseous-dusty astronomical media (Largo 2004, Woon 2002, and
references therein). Despite the success of ab initio theoretical
calculations, the endothermic ion-molecule reactions have always
been discounted, and only exothermic reactions have been accepted as
a viable mechanism. However, with the knowledge of the kinetic
energy (or at least with its value range) of some radical and ionic
fragments, some endothermic ion-molecule reactions could be
competitive and, in extreme situations, even become more efficient
than those exothermic reactions.

We have determined the kinetic energy of all cationic fragments from
the photodissociation of formic acid.

The present time-of flight spectrometer was designed to fulfil the
Wiley-McLaren conditions for space focusing (Wiley \& McLaren 1955).
Within the space focusing conditions, the observed broadening of
peaks in spectra is mainly due to kinetic energy release of
fragments. Considering that the electric field in the interaction
region is uniform, we can determine the released energy in the
fragmentation process ($U_0$) from each peak width used by Simon et
al. (1991), Hansen et al. (1998), Santos, Lucas \& de Souza (2001)

\begin{equation} \label{eq-U0}
U_0 = \Big(\frac{qE \Delta t}{2} \Big)^2 \frac{1}{2m}
\end{equation}
where $q$ is the ion fragment charge, $E$ the electric field in
interaction region, $m$ is the mass of fragment, and $\Delta t$ is
the time peak width (FWHM) taken from PEPICO spectra. In order to
test the above equation we have measured the argon mass spectrum
under the same conditions. The kinetic energy value achieved for the
Ar$^+$ ions is in agreement with the mean kinetic energy $(3/2) KT$
obtained assuming Maxwell's distribution law. The calculated values
for kinetic energy release ($U_0$) for formic acid fragmentation is
shown in Table~\ref{tab-piy}.

We observe that the highest kinetic energy release was associated
with the lightest fragment H$^+$ ($m/q=1$), as expected. Extremely
fast ionic fragments ($U_0 > 10$ eV), usually associated with
dissociation of doubly or multiply-charged ions were also observed
at high photon energies. These observations point to the important
role of Auger process ion fragmentation of core-ionized polyatomic
molecules. The Coulomb explosion associated with the Auger process
should explain the increase in kinetic energy of the ionic
fragments, reflected in the increasing broadening of several
fragments. This broadening observed in simple coincidence spectra
(PEPICO) and its consequence on the shape of peaks in mass spectra
has been discussed by Simon et al. (1993).

Averaging the kinetic energy release $U_0$ of each fragment by the
fragmentation PIY we have found the mean kinetic energy release
$<U_0>_\varepsilon$ due to a photon of energy $\varepsilon$,
considering all fragments in single coincidence spectra.

\begin{equation} \label{eq-U0}
<U_0>_\varepsilon = \frac{\sum_f PIY_f \times U_{0_f}} {100}
\end{equation}

\begin{figure}
 \resizebox{\hsize}{!}{\includegraphics{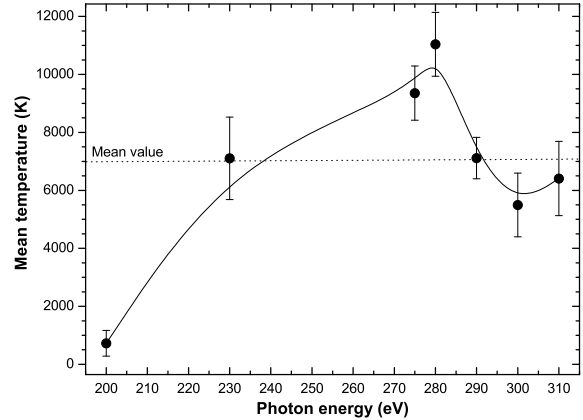}}
\caption{Mean Kinetic temperature of ionic fragments of formic acid
molecule as a function of photon energy.} \label{fig-heating}
\end{figure}

Assuming Maxwell's distribution of velocities we calculated the mean
kinetic temperature of all fragments in the interaction region. This
temperature can be seen in Figure~\ref{fig-heating} as a function of
photon energy. Photons with energies close to the C 1s threshold
photoionization energy seem to produce faster ionized fragments
which can been associated with some kind of resonance in molecule.
Averaging over all photon energies we can estimate the total heating
$H$ of ionic fragmentation release above room temperature by

\begin{equation} \label{eq-U0}
H = <T_0> - T_{room} = \frac{2}{3} \frac{1}{k}
\frac{\sum_{\varepsilon} <U_0>_\varepsilon }{n} - 290 K
\end{equation}
where $<T_0>$ is the mean kinetic temperature of all cationic
fragments averaged in all photon energies, k is the Boltzmann
constant ($8.61 \times 10^{-5} $ eV K$^{-1}$, $\varepsilon$ is the
photon energy in eV and n is the total number of spectra in our
experimental data.

We have found the value $H \sim 7000 K$ ($\sim$ 0.9 eV). This result
shows that photoionization by an X-ray field can increase the local
temperature since several fragments could reach high values of
kinetic energy.

As the C 1s ionizing potential of formic acid is 295.8 eV (Prince et
al. 2003) and using soft X-ray photons, the heating $H$ produced by
ejected inner shell photoelectrons is about 5-15 times greater than
that caused by ionic fragments.

The study of the decay of core-excited molecules provides
information about the bonding or antibonding nature of the molecular
orbitals. Generally, the final electronic states of a core excited
molecule are unknown due to the fact that the densities of the
states are very high, and the bond distances and angles differ from
their ground state configuration. The surface potentials of the
ionic states are extremely repulsive. For core excited molecules
which dissociate in one charged and one or more neutral fragments,
the dissociation is primarily controlled by chemical (non-Coulomb)
forces originating from the residual valence electrons of the system
(Nenner \& Morin, 1996). From Table 1, one can see that the mean
kinetic energy release $U_0$ of some formic acid fragments increases
as the photon energy approximates the C 1s edge (288 eV). This
enhancement is due to the repulsive character of the $\sigma$*
($\pi$*) resonance.

The maximum possible kinetic energy release available to a molecular
fragment is given by the difference between the deposited photon
energy and the appearance energy of the fragment. Lorquet (2000) has
shown that the relationship between the average kinetic energy
release $U_0$ and the internal energy measured in excess of the
dissociation threshold, $\Delta E$, is not linear. It contains
information about the density of vibrational-rotational states. The
mean kinetic energy release is found to increase as $\Delta
E^{1/2}$. This behavior can be seen in Figure~\ref{fig-heating}.

\subsection{Absolute photoionization and photodissociation cross sections}

Sorrell (2001) proposed a theoretical model for the chemical
manufacture of interstellar aminoacids and sugars, assuming that
those biomolecules are formed inside the bulk interior of icy grain
mantles photoprocessed by starlight (ultraviolet and soft X-rays
photons). His model chemistry is based on radical-radical reactions
followed by chemical explosions of a processed mantle that ejects
large amounts of organic dust into the ambient gaseous medium. The
density number of a given biomolecule in a steady state regime of
creation and destruction inside a gaseous-dusty cloud is given by
\begin{equation}
N_{Mol} = \frac{\dot{N}_{Mol} n_d}{<\sigma_{ph-d}> I_{0}}
\end{equation}
where $\dot{N}_{Mol}$ is the molecule ejection rate which depends
mainly on the molecule mass and the properties of grains and the
cloud (see eq. 21 of Sorrell 2001), $n_d$ is the dust space density
and $I_{0}$ is the number flux of ionizing photons (photons
cm$^{-2}$ s$^{-1}$) inside the cloud and $<\sigma_{ph-d}>$ is the
averaged photodissociation cross section in wavelength range of the
photon flux density.

As mentioned by Sorrell (2001), the main uncertainty of this model
equilibrium abundance cames from the uncertainty of $\sigma_{ph-d}$
value. Therefore the precise determination of $\sigma_{ph-d}$ of
biomolecules is very important to estimate the molecular abundance
of those molecules in the interstellar medium. Moreover, knowing the
photon dose $\phi$ and $\sigma_{ph-d}$ value we can also determine
the half-life of given molecule as discussed by Bernstein et al.
(2004).

The photodissociation rates, $R$, of a molecule dissociated by the
interstellar radiation field $I_\varepsilon$ in the energy range,
$\varepsilon_2 - \varepsilon_1$, is given by
\begin{equation} \label{eq-R}
R = \int_{\varepsilon_1 }^{\varepsilon_2} \sigma_{ph-d}(\varepsilon)
I_0(\varepsilon) d\varepsilon
\end{equation}
where $\sigma_{ph-d}(\varepsilon)$ is the photo-dissociation cross
section as a function of photon energy (cm$^2$), $I_0(\varepsilon)$
is the photon flux as a function of energy (photons cm$^{-2} eV^{-1}
s^{-1}$) (see discussion in Cottin et al. 2003 and Lee 1984).

From Eq.~\ref{eq-R} we can also derive the half-life time,
$t_{1/2}$, of the molecule as
\begin{equation}
t_{1/2} = \frac{\ln 2}{\int_{\varepsilon_1 }^{\varepsilon_2}
\sigma_{ph-d}(\varepsilon) I_0(\varepsilon) d\varepsilon}
\end{equation}
which does not depend on the molecular number density.

In order to put our data on an absolute scale, we have summed all
the contributions of all cationic fragments detected and normalized
them to the photoabsorption cross sections measured by Ishii \&
Hitchcock (1987).

\begin{figure}
 \resizebox{\hsize}{!}{\includegraphics{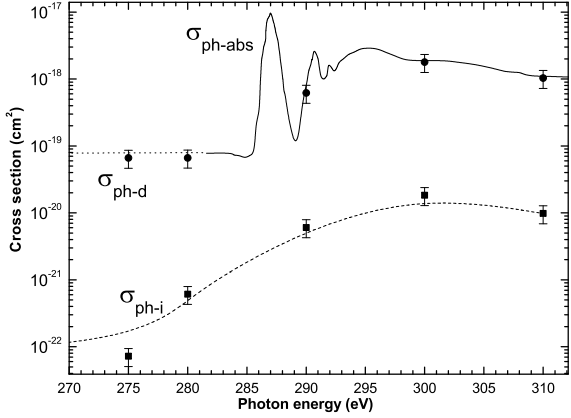}}
\caption{Non-dissociative single ionization (photoionization) cross
section ($\sigma_{ph-i}$) and dissociative ionization
(photodissociation) cross section ($\sigma_{ph-d}$) of formic acid
as a function of photon energy. The photoabsorption cross section
$\sigma_{ph-abs}$ taken from Ishii \& Hitchcock (1987) is also
shown.}
 \label{fig-sigma}
\end{figure}

After a subtraction of a linear background and false coincidences
coming from aborted double and triple ionization (see Simon et al.
1991), the integrated true coincidence signal $I_f^{q+}$ for a given
fragment $f$ with charge $q$, at constant photon flux and lower
target pressures is given by
\begin{equation} \label{eq-1}
I_f^{q+} = \epsilon_{e_{q+}} \epsilon_i I_{0} n t_{exp}
\sigma_f^{q+} = \epsilon_{e_{q+}} \epsilon_i K \sigma_f^{q+}
\end{equation}
where $I_{0}$ is the total photon number that reaches the molecular
beam (photons s$^{-1}$), $n$ is the number of target scattering
centers per area (as function of the collision region pressure),
$t_{exp}$ is the experiment acquisition time, $\sigma^{q+}$ is the
ionization cross section \footnote{$\sigma^{+}$, $\sigma^{2+}$ and
$\sigma^{3+}$ are the single, double and triple ionization cross
section obtained form PEPICO, PE2PICO and PE3PICO spectra,
respectively.} $\epsilon_i$ and $\epsilon_{e_{q+}}$ are the ion and
the electron \footnote {The $\epsilon_{e_{2+}}$ is the efficiency of
detecting at least one of two electrons released by double
ionization processes in PE2PICO spectra and $\epsilon_{e_{3+}}$ is
the efficiency of detecting at least one of three electrons released
by triple ionization processes in PE3PICO spectra} detection
efficiencies, respectively, where once again $q$ designates the
number of ejected electrons. The adopted efficiency values were
$\epsilon_{i} = 0.23$ ; $\epsilon_{e_{1+}} = 0.04$;
$\epsilon_{e_{2+}} = 0.54$; $\epsilon_{e_{3+}} = 0.78$ (Cardoso
2001).

Since the absolute target density $n$ as well as the incident number
of photons $I_0$ have not been determined absolutely, we take $K
\equiv  I_0 n t_{exp} $ as a constant (in cm$^{-2}$) to determined
for each given energy photon .

The total integrated coincidence signal for a given photon energy
and final charge state $q+$ is given by
\begin{equation}
I^{q+}=\sum_f I_f^{q+} = \epsilon_{e_{q+}} \epsilon_i^q K
\sigma^{q+}
\end{equation}

Analogously, the total ionization cross section is given by
\begin{equation} \label{eq-2}
\sigma^{q+}= \sum_f \sigma_f^{q+} = \frac{1}{K}
\frac{I^{q+}}{\epsilon_{e_{q+}} \epsilon_i^q}
\end{equation}

Assuming a negligible fluorescence yield (due to the low carbon
atomic number (Chen et al. 1981)) and anionic fragments production,
we adopted that all absorbed photon lead to cationic ionizing
process, therefore we can write
\begin{equation}
\sigma_{abs}\equiv \sum_{q=1}^3  \sigma^{q+} =
\frac{1}{K}\sum_{q=1}^3  \frac{I^{q+}} {\epsilon_{e_{q+}}
\epsilon_i^q}
\end{equation}
or yet

\begin{table}
\centering \caption{Values of non-dissociative single ionization
(photoionization) cross section ($\sigma_{ph-i}$) and dissociative
ionization (photodissociation) cross section ($\sigma_{ph-d}$) of
formic acid as a function of photon energy. The estimated
experimental error was 30\%. The photoabsorption cross section
$\sigma_{ph-abs}$ from Ishii \& Hitchcock 1987.} \label{tab-sigma}
\begin{tabular}{ l l c c c }
\hline \hline
Photon       &  & \multicolumn{3}{c}{Cross Sections (cm$^{2}$)}\\
\cline{3-5} energy (eV)  &  & $\sigma_{ph-d}$ & $\sigma_{ph-i}$ &
$\sigma_{ph-abs}$
\\
\hline
275          &  & 6.61 $\times 10^{-20}$  & 7.21 $\times 10^{-23}$  & 7.00$^a$ $\times 10^{-20}$ \\
280          &  & 6.64 $\times 10^{-20}$  & 6.12 $\times 10^{-22}$  & 7.00$^a$ $\times 10^{-20}$ \\
290          &  & 6.20 $\times 10^{-19}$  & 6.06  $\times 10^{-21}$ & 6.42 $\times 10^{-19}$ \\
300          &  & 1.78 $\times 10^{-18}$  & 1.83 $\times 10^{-20}$  & 1.88 $\times 10^{-18}$ \\
310          &  & 1.03 $\times 10^{-18}$  & 9.83 $\times 10^{-21}$  & 1.10 $\times 10^{-18}$ \\
\hline \hline
\multicolumn{5}{l}{$^a$ Extrapolated data (see Fig~\ref{fig-sigma})}\\
\end{tabular}
\end{table}

\begin{equation} \label{eq-4}
K = \frac{\sum_{q=1}^3 \frac{I^{q+}}{\epsilon_{e_{q+}}
\epsilon_i^q}}{\sigma_{abs}}
\end{equation}

Therefore, we can calculate the cross section for single ionized
($q=1$) fragments:

\begin{equation}
\sigma^{+}=  \frac{I^{+}} {\epsilon_{e_{+}} \epsilon_i} \frac{1}{K}
\end{equation}

The non-dissociative single ionization (photoionization) cross
section $\sigma_{ph-i}$ and the dissociative single ionization
(photodissociation) cross section $\sigma_{ph-d}$ of formic acid can
be determined by
\begin{equation}
\sigma_{ph-i} = \sigma^{+} \frac{PIY_{HCOOH^+}}{100}
\end{equation}

In the same way the photodissociation cross section of formic acid
can be determined by
\begin{equation}
\sigma_{ph-d} = \sigma^{+} \Big( 1 - \frac{PIY_{HCOOH^+}}{100} \Big)
\end{equation}

Both cross sections can be seen in Figure~\ref{fig-sigma} as a
function of photon energy. The absolute absorption cross section of
formic acid (Ishii \& Hitchcoch 1987) is also shown for comparison.
Those values are also shown in Table~\ref{tab-sigma}.

The cross sections determined here should be useful in discussions
of interstellar chemistry.

\section{Summary and conclusions}

The goal of this work is to experimentally study ionization and
photodissociation processes of a glycine precursor molecule, HCOOH
(formic acid). The measurements were taken at the Brazilian
Synchrotron Light Laboratory (LNLS), employing soft X-ray photons
from a toroidal grating monochromator TGM) beamline (200 - 310 eV).
The experimental set-up consists of a high vacuum chamber with a
time-of-flight mass spectrometer TOF -MS. Mass spectra were obtained
using PhotoElectron PhotoIon Coincidence (PEPICO) technique. Kinetic
energy distributions and abundances for each ionic fragment were
obtained from the analysis of the corresponding peak shapes in the
mass spectra.

We have shown that the X-ray field interactions with formic acid
release a considerable number of energetic fragments, some of then
with high kinetic energy (ex. H$^+$, CH$^+$, O$^+$) and some with
extreme high kinetic energy (CO$^{+++}$, CO$^{++}$, O$^{++}$). Those
energetic fragments promote an increase in the local temperature of
a region. An extension of this scenario to interstellar medium
conditions could justify (or promote) some endothermic ion-molecule
reactions and become important in elucidating some pathways of
formation of complex molecules (Largo et al. 2004).

Dissociative and non-dissociative photoionization cross sections
were also determined. Due to the high photodissociation cross
section of formic acid it is possible that in PDR regions, just
after molecule evaporation from the grain surface, it is almost
destroyed by soft X-rays, justifying the observed low abundance of
HCOOH in the gaseous phase. This result could indicate that the
preferential path for the glycine formation via formic acid may be
through ice.

%
\begin{acknowledgements}The authors would like to thank the staff
of the Brazilian Synchrotron Facility (LNLS) for their valuable help
during course of the experiments. We are particularly grateful to
Professor G.G.B. de Souza and Professor A. N. de Brito for the use
of Time-of-Flight Mass Spectrometer. This work is supported by LNLS,
CNPq and FAPERJ.
\end{acknowledgements}
%
%

\end{document}